# Can a Mobile Game Teach Computer Users to Thwart Phishing Attacks?

Nalin Asanka Gamagedara Arachchilage[1]
Cyber Security Centre
Department of Computer Science
University of Oxford
OX1 3QD, UK

Steve Love[2]
School of Information Systems Computing and Mathematics
Brunel University
Uxbridge, Middlesex
UB8 3PH, UK

Carsten Maple[3]
Institute for Research in Applicable Computing
University of Bedfordshire
Luton, Bedfordshire
LU1 3JU, UK

## Abstract

*Phishing is an online fraudulent technique, which aims to steal sensitive information such as usernames, passwords and online banking details from its victims. To prevent this, anti-phishing education needs to be considered. This research focuses on examining the effectiveness of mobile game based learning compared to traditional online learning to thwart phishing threats. Therefore, a mobile game prototype was developed based on the design introduced by Arachchilage and Cole [3]. The game design aimed to enhance avoidance behaviour through motivation to thwart phishing threats. A website developed by Anti-Phishing Work Group (APWG) for the public Anti-phishing education initiative was used as a traditional web based learning source. A think-aloud experiment along with a pre- and post-test was conducted through a user study. The study findings revealed that the participants who played the mobile game were better able to identify fraudulent web sites compared to the participants who read the website without any training.*

## 1. Introduction

Internet technology is so pervasive today that it provides the backbone for modern living enabling people to shop, socialize, communicate and be entertained all thorough their personal computers connected to the Internet. As people's reliance on the Internet grows, so the possibility of hacking and other security breaches increases rapidly [13]. This is mainly because sensitive trust decisions are made during online activities; such as online banking transactions or bill payments. Therefore, professionalism, training and education are worth considering in order protecting people from cyber-attacks.

Cyber-attacks can include malicious IT threats such as a set of computer programs that can disturb the normal behaviour of computer systems (viruses), malicious software (malware), unsolicited e-mail (spam), monitoring software (spyware), attempting to make computer resources unavailable to its intended users (Distributed Denial-of-Service or DDoS attack), the art of human hacking (social engineering) and online identity theft (phishing). The motivation behind these attacks tends to be for, either financial or social gain [14, 25 and 26]. For example, a DDoS attack could target a bank in order to break down their email server and the attacker can exhort a lump sum of money to give the email server back to the bank.

However, a cyber-threat that is particularly dangerous to computer users is phishing. Phishing is a form of semantic attack [7 and 21], that leverages human vulnerabilities, rather than exploiting technical pitfalls. In phishing, victims get invited by scam emails to visit fraudulent websites. The attacker creates a mimic website which has the look-and-feel of the legitimate website. Innocent users are invited by sending emails to access to the mimic website and steal their money. Phishing attacks get more sophisticated day by day as and when attackers learn new techniques and change their strategies accordingly [10 and 11].

A number of automated anti-phishing tools have been developed and used to alert users of potentially fraudulent emails and websites. For example, Calling ID Toolbar, Cloudmark Anti-Fraud Toolbar, EarthLink Toolbar, Firefox 2, eBay Toolbar and Netcraft Anti-Phishing Toolbar. Ye and Sean [27] and Dhamija and Tygar [6] have developed a prototype called "trusted paths" for the Mozilla web browser that was designed to help users verify that their browser has made a secure connection to a trusted website. However, these tools are not entirely reliable in combating phishing threats [22 and 4]. Zhang, et al. [28] has reported that even the best anti-phishing tools missed over 20 percent of phishing websites.

In relation to this, however, research has also revealed that well designed end-user security education can be effective [9, 10, 11 and 23]. This could be web-based training materials, contextual training and embedded training to enhance users' ability to avoid phishing threats. One objective of the current research is to find effective ways to educate people how to detect and prevent from phishing attacks.

So, how does one can educate computer users in order to protect them from becoming the victims of phishing attacks? The study reported in this paper attempts to evaluate the effectiveness of mobile game based learning compared to traditional web based learning to thwart phishing threats. This concept is grounded on the notion that not only mobile games can provide education [20], but also games potentially offer a better natural learning environment that motivates the user to keep engaging with it [1 and 18]. In addition, game based education attracts and retains the user till the end of game by providing immediate feedback or response.

The most significant feature of a mobile environment is "mobility" itself such as mobility of the user, mobility of the device and mobility of the service [15]. It enables users to be in contact while they are outside the reach of traditional communicational spaces. For example, a person can play a game on his mobile device while travelling on the bus or train, or waiting in a queue.

## 2. Methodology

To accomplish this research study, a mobile game prototype was developed using MIT App Inventor emulator based on the design introduced by Arachchilage and Cole [3]. The overall mobile game prototype was designed to enhance the user's avoidance behaviour through motivation to protect themselves against phishing attacks. Additionally, a website developed by Anti-Phishing Work Group (APWG) for the purpose of public anti-phishing education initiative was employed as a traditional web based learning source in this research study.

### 2.1. Mobile game design prototype

Arachchilage and Cole [3] designed a mobile game prototype as an educational tool to teach computer users how protect themselves against phishing threats. The research study asked the following questions: The first question is how does one identify which issues the game needs to be addressed? Once the salient issues were identified, the second question is what principles should guide the structure of this information. The elements from a theoretical model derived from Technology Threat Avoidance Theory (TTAT) were used to address those mobile game design issues and the mobile game design principles were used as a set of guidelines for structuring and presenting information in the mobile game design context [3 and 13]. The objective of their anti-phishing mobile game design was to teach the user how to identify phishing URLs (Uniform Resource Locator). The overall mobile game design was focused to enhance avoidance behaviour through motivation of computer users to thwart phishing threats. The prototype game design was presented on MIT App Inventor Emulator as shown in Figure 1.

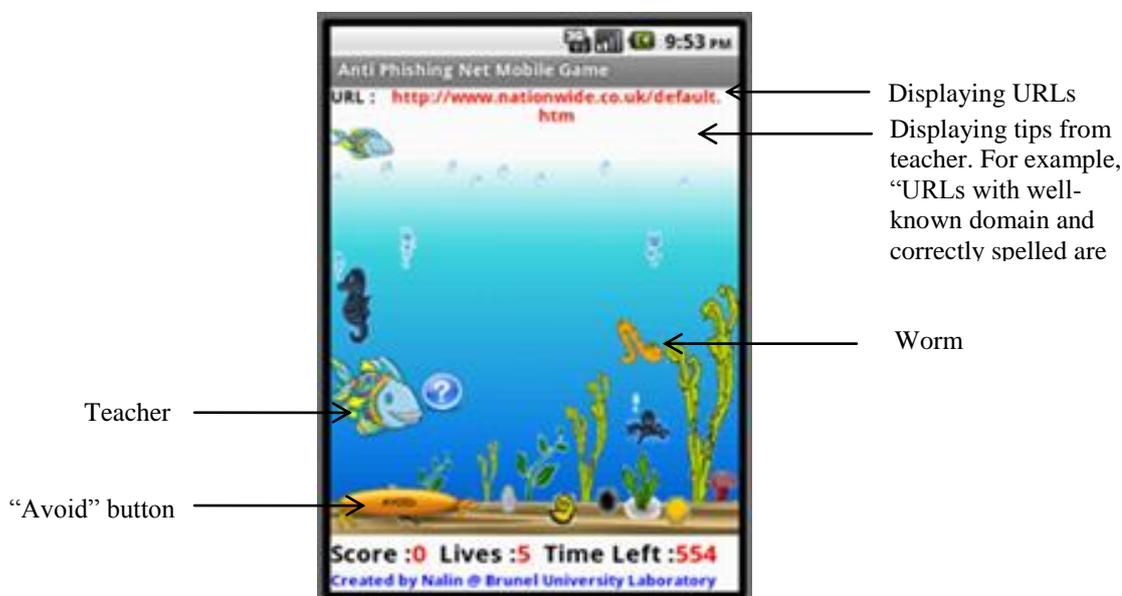

Figure 1: The mobile game prototype on MIT App Inventor Emulator

A URL is displayed with each worm where the worms are randomly generated. If the worm associated with URL is legitimate, then the user is expected to tap on the worm in order to increase the score. However, if the user fails to identify the legitimate URL, then remaining lives will be reduced by one point. On the other hand, if the worm associated with the URL is phishing, then the user is also expected to tap on "AVOID" button to reject the URL in order to increase the score. If the user fails to do this, then remaining lives will be reduced by one point. If the worm associated with the URL is suspicious and if it is difficult to identify, the user can tap on big fish (in this case, teacher fish) to request help. Then some relevant tips will be displayed just below the URL. For example, "website addresses associate with numbers in the front are generally scams. Whenever the user taps on the big fish, the time left will be reduced by 100 points (in this case 100 seconds). Finally, the user will gain 10 points if all the given URLs were correctly identifed within 5 lives and 600 seconds to complete the game.

## 2.1. Mobile game design prototype

The Anti-Phishing Work Group (APWG) was established in 2003 as an industry association focused on amalgamating the global response to cyber-crime [2]. The organization provides a forum for public such as responders and managers of cyber-crime to discuss phishing and other cyber-crime issues, to consider potential technology solutions, to access data logistics resources for cyber-security applications and for cybercrime forensics, to cultivate the university research community dedicated to cyber-crime and to advise government, industry, law enforcement and treaty organizations on the nature of cybercrime.

The public phishing education section of the Anti-Phishing Work Group website is developed for learning more about phishing education. For example, what is phishing threat, how it could be severe, what is the usefulness of having a safeguarding measure, where to report a suspected phishing email or website and phishing education to thwart phishing attacks. Therefore, the public Anti-phishing education initiative section of the APWG website was used as a traditional online learning source in our research study, which is shown in Figure 2.

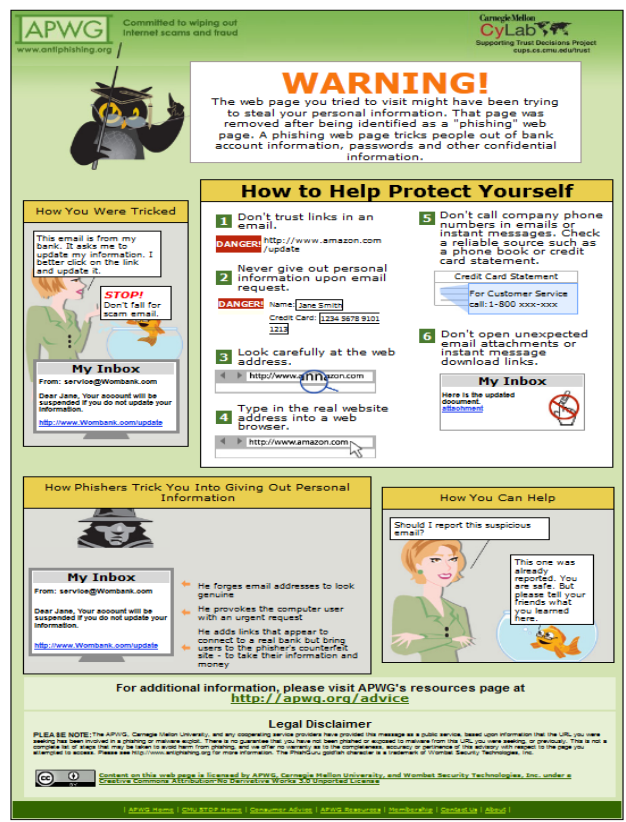

Figure 2: Educational redirect program section of APWG Public Education [2]

### 2.3 Participants

Sheng, et al. [22] have conducted a role-play survey with 1001 online survey respondents to study who falls for phishing attacks. Their study showed participants between the ages of 18 to 25 are more susceptible for phishing attacks than other age groups. The study reported in this paper included 40 participants from a diverse group of staff and student at Brunel University and the University of Bedfordshire, including people who were concerned about computer security. Participants' ages ranged from 18 to 25, with a gender split of 67 per cent male and 33 per cent female. They had average of 16 -20 hours per week of Internet experience (SD=1.19). Each participant took part in the think-aloud study on a fully voluntary basis. A summary of the demographics of the participants in the think-aloud study is shown in Table 4.

Table 1: Participant demographics

| Characteristics | Total (Mobile game prototype) | Total (APWG educational website) |
|---|---|---|
| Sample Size | 20 | 20 |
| Gender<br>  Male<br>  Female | <br>13<br>7 | <br>13<br>7 |
| Age (18 - 25) | 20 | 20 |
| Experience using mobile device<br>  Mobile phone<br>  Smart phone | <br><br>0<br>20 | <br><br>0<br>20 |
| Average hours per week on the Internet<br>  0-5<br>  6-10<br>  11-15<br>  16-20<br>  20+ | <br><br>0<br>0<br>0<br>0<br>20 | <br><br>0<br>0<br>0<br>0<br>20 |

## 2.4 Procedure

The pre- and post-tests were based on Apple MacBook Pro where the participants received their score at the end of each test. First and foremost, each individual participant was explained the nature of the think-aloud experimental study and asked to sign a consent form. They were also informed that the experiment is about testing their understanding of phishing threat awareness using either mobile game prototype or APWG public education initiative website. Then the individual participants were asked whether or not they knew what the term 'phishing attack' means. Those who gave a positive response, were asked to give a short verbal description to confirm their understanding, whilst negative responders were read a brief definition of phishing attack and gave a short verbal description. To begin the experiment, total 40 participants were asked to follow think-aloud user study instructions given in an experimental protocol. Participants were randomly assigned to two groups: 20 participants with a group those who played the mobile game and the other 20 participants with another group those who read the APWG public education initiative website. They were also informed that they are welcome to clarify anything related to the experiment. In the pre-test, participants were presented with ten websites and asked to differentiate phishing websites from legitimate ones. After evaluating 10 websites (Table 2), participants were given fifteen minutes to complete a mobile game based training activity on a HTC One X touch screen smart phone. Initially, the game was designed with 10 suspect URLs where the participant's responsibility is to identify legitimate URLs from phishing ones.

This research study employed a tool called System Usability Scale (SUS), which is used to measure users' subjective satisfaction of mobile game interface usability since we developed the mobile game prototype. Brooke [5] stated that the SUS is generally used after the respondent has had an opportunity to use the system (in this case the mobile game or website) being evaluated, however before any debriefing or discussion takes place. Furthermore, he stated that the conditions of the study, sample sizes of at least 12-14 participants are needed to get reasonably reliable results. The SUS uses a five-point Likert scale with anchors for strongly agree and strongly disagree. Therefore, after engaging 15 minutes with the mobile game activity, 20 participants were asked to fill in a survey (SUS questionnaire items), which was used to measure the participant's subjective satisfaction of the mobile game prototype interface. The other 20 participants were asked to walkthrough the APWG public education initiative website for 15 minutes.

A total of 40 participants were followed by a post-test where participants were shown ten more websites to evaluate (Table 3). The score was recorded during the pre- and post-tests to observe how participants' understanding and the awareness of phishing threats developed through the mobile game based learning. More than half of the websites were phishing websites based on popular brands, whilst the rest were legitimate websites from popular brands. For the purpose of this test, recently being attacked phishing websites were captured from PhishTank.com [17] from November 1 to November 28, 2012. All phishing website URLs were selected within 7 hours of being reported. During the mobile game based training and public education initiative website reading activities, a think-aloud study was employed where participants talked about their opinions and experience of phishing threat awareness and understanding through either mobile game prototype or APWG public education initiative website.

Table 2: List of ten website addresses used in pre-test

| Real or Phishing | Website Name | Website address |
|---|---|---|
| Phishing | Santander | 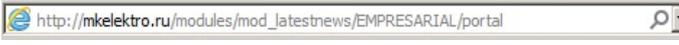 |
| Phishing | PayPal | 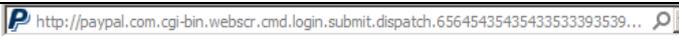 |
| Real | HSBC | 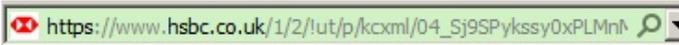 |
| Phishing | Halifax | 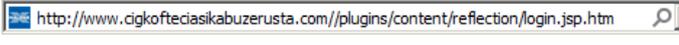 |
| Real | eBay | 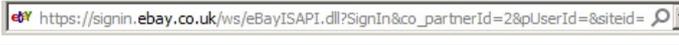 |
| Phishing | Western Union | 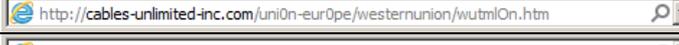 |
| Phishing | eBay | 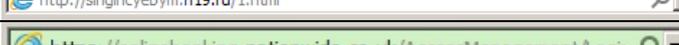 |
| Real | Nationwide | 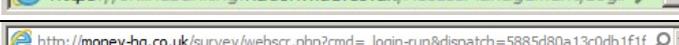 |
| Phishing | Money:hq | 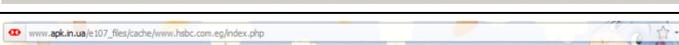 |
| Phishing | HSBC | 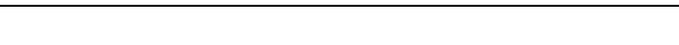 |

Table 3: List of ten website addresses used in post-test

| Real or Phishing | Website Name | Website address |
|---|---|---|
| Phishing | PayPal | 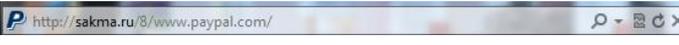 |
| Phishing | HABBO | 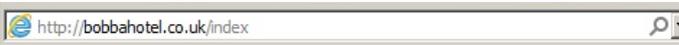 |
| Real | FDIC | 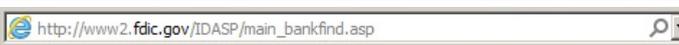 |
| Phishing | Littlewoods | 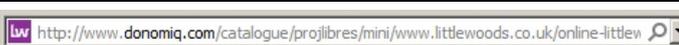 |
| Real | Lloyds TSB | 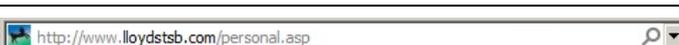 |
| Phishing | Facebook | 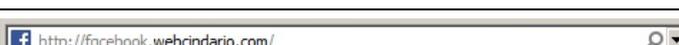 |
| Phishing | Santander | 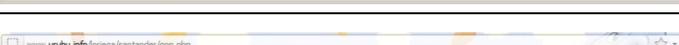 |
| Real | UPS | 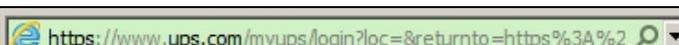 |
| Phishing | eBay | 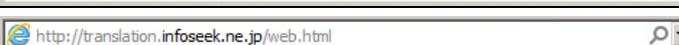 |
| Phishing | AOL | 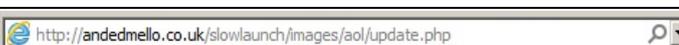 |

## 3. Results

In the think-aloud experiment, participants talked about their experience and opinions of either mobile game prototype or APWG public education initiative websiteto thwart phishing threats. The results were encouraging; however it highlighted some areas where the APWG public education initiative website needed improvements.

Initially, we evaluated the participants' subjective satisfaction of the mobile game prototype (since we developed the prototype) using SUS scoring approach introduced by Brooke [5]. The score was significantly high with 84 percent (83.62 out of 100) [5]. Then the research study employed ***Paired-samples t-test*** to compare the means scores for the participants' pre- and post-tests [16]. Participants who played the mobile game, scored 56 percent in the pre-test and 84percent in the post-test of identifying phishing or legitimate websites after playing the mobile game prototype (Table 4 and 5). There was a statistically significant increase in the post-test of participants who played the mobile game ((Pre-test: M= 55.00, SD=17.911 and Post-test: M=84.00, SD=13.139), t(19)= -7.97, p<0.005 (two-tailed)) compared to those who read APWG public education initiative website (Pre-test: M= 60.00, SD= 17.770 and Post-test: M= 62.50, SD=25.930, t(19)= -0.036, p>0.005 (two-tailed)).

It has been seen that there is a considerable improvement of participants' results of 29 percent in the post-test after their engagement with the mobile game prototype (p<0.005 (two-tailed)). This is a significant improvement of overall participants' phishing avoidance behaviour through the mobile game. 18 participants scored above 80 percent whilst five participants scored full marks (100 percent) in the post-test. However, all participants scored above 50 percent in their post-test. The overall participants' score is shown in Figure 3. Additionally, the individual participant's score during their engagement with the mobile game prototype is shown in Figure 4.

Table 4: Paired Samples t-test Statistics

| | | Mean | N | Std. Deviation | Std. Error Mean |
|---|---|---|---|---|---|
| Pair 1 | MobilePreTest | 55.00 | 20 | 17.911 | 4.005 |
| | MobilePostTest | 84.00 | 20 | 13.139 | 2.938 |
| Pair 2 | WebPreTest | 60.00 | 20 | 17.770 | 3.974 |
| | WebPostTest | 62.50 | 20 | 25.930 | 5.798 |

Table 5: Paired Samples Test

| | | Paired Differences | | | | | t | df | Sig. (2-tailed) |
|---|---|---|---|---|---|---|---|---|---|
| | | Mean | Std. Deviation | Std. Error Mean | 95% Confidence Interval of the Difference | | | | |
| | | | | | Lower | Upper | | | |
| Pair 1 | MobilePreT - MobilePostT | -28.500 | 15.985 | 3.574 | -35.981 | -21.019 | -7.973 | 19 | .000 |
| Pair 2 | WebPreT - WebPostT | -2.500 | 31.267 | 6.992 | -17.133 | 12.133 | -.358 | 19 | .725 |

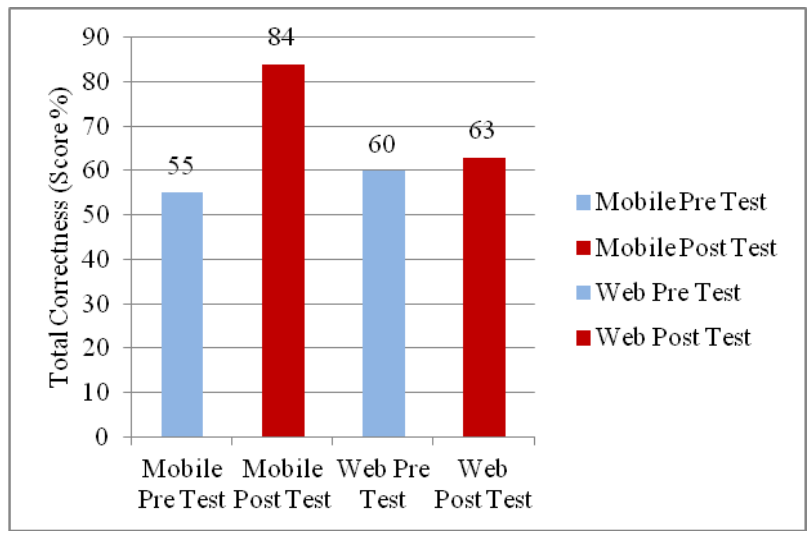

Figure 3: Total correctness for the test groups. N=20 in all conditions. The game condition shows the greatest improvement (by 29%).

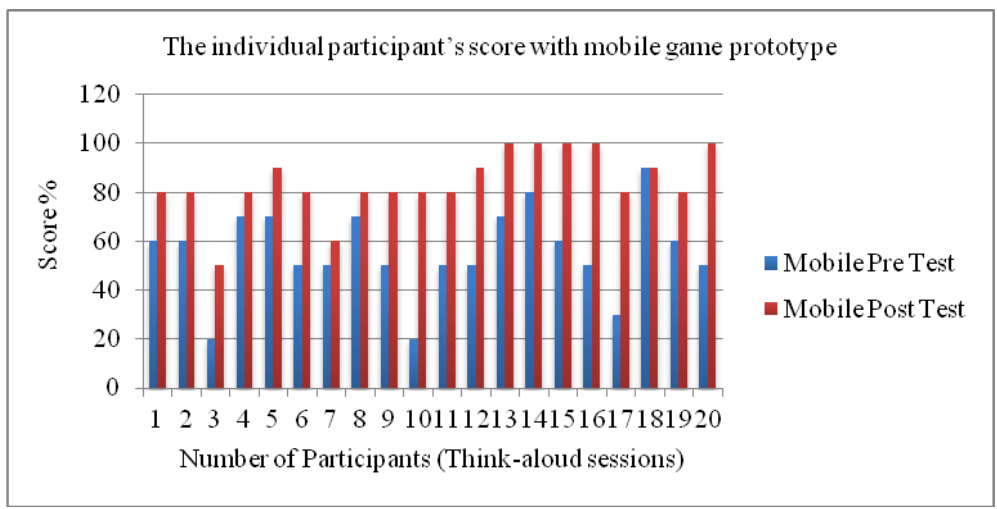

Figure 4: The individual participant's score during their engagement with the mobile game prototype.

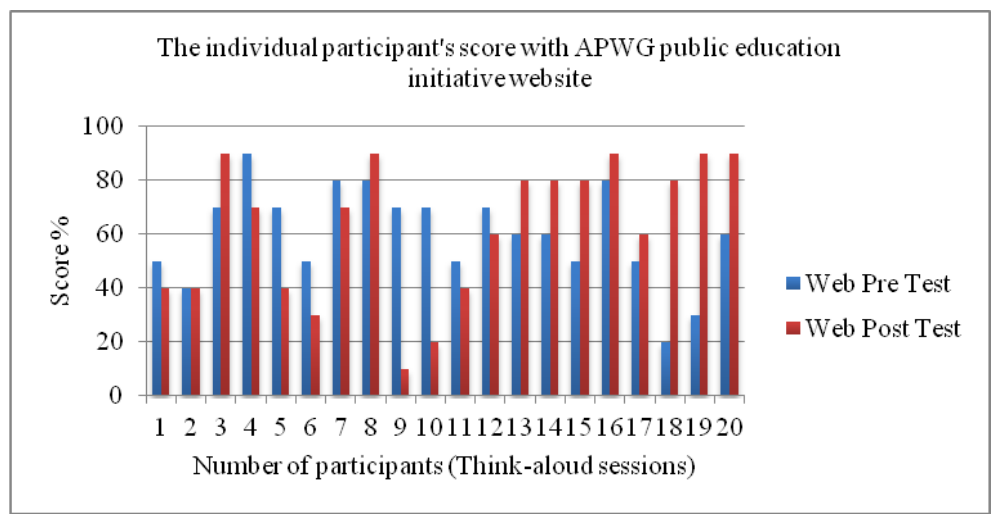

Figure 5: The individual participant's score during their engagement with the APWG public education initiative website

All participants responded that their decision was based on looking at the address bar (URL) when evaluating the second set of ten websites in the post-test. Furthermore, they stated that they made only very few attempts to look at the address bar when evaluating the first ten websites in the pre-test. There, many participants used incorrect strategies to determine the website legitimacy. For example, one of the common strategies consisted of checking whether or not the website was designed professionally. However, this may not be a useful strategy as many phishing websites are exact replica of legitimate websites. The attacker can easily mimic any professional website from the source code of the particular page provided by the browser. Moreover, all participants highlighted that the mobile game was somewhat effective at teaching different patterns of URLs to differentiate phishing URLs from legitimate ones.

Participants those who read the APWG public education initiative website, scored 60 percent in the pre-test and 63 percent in the post-test of identifying phishing or legitimate websites after reading the APWG public education initiative website (Table 4 and 5). There is a slightly little improvement of participants' results of 3 percent in the post-test during the think-aloud study. 7 participants scored below 50 percent whilst five participants scored above 80 percent in the post-test. Only 12 participants scored above 50 percent in the post-test. None of the participants scored full marks (100 percent) in their post-test. The overall participants' score is shown in Figure 3. Additionally, the individual participant's score during their engagement with the APWG public education initiative website is shown in Figure 5. It can therefore be argued that participants were not much aware of how to avoid phishing threats by just reading the APWG public education initiative website. The current research study identified possible explanations to these results. All participants who read the website during the think-aloud study stated that they are bored of reading the website, since it contains a lot of text. Their argument was reading a website to gain phishing knowledge does consume a lot of time, though it is useful in the long run. Only a few participants responded that their decision was based on looking at the address bar when evaluating the second set of ten websites in the post-test.

Furthermore, they described that the website does tell us to look at the URL; however, it does not tell us how to identify good URLs from bad ones more specifically. They also believed that the website does not provide a way of testing their knowledge. For example, it would be useful to give the user a test after reading the website for self-assessment their knowledge gained against phishing attacks. This helps to obtain an immediate feedback on what they have learnt. Additionally, video tutorials can be embedded into the website where the user can better engage with learning process. Therefore, the current study concluded that the mobile game prototype is somewhat effective in teaching computer users to thwart phishing attacks than traditional online learning.

## 4. Implications for Research

Furthermore, the research work reported in this paper focuses on how one can educate computer users against phishing threats, because phishing attacks continuously jeopardize innocent computer users [8, 9 and 19]. Usability experts claim that end-user education and training does not work to combat against phishing [23 and 29]. It can therefore be argued that existing anti-phishing security education such as through web-based training, contextual training and embedded training materials has not been appropriately designed for users. Yet theory-based empirical research that enhances computer users' phishing awareness through their voluntary IT threat avoidance behaviour is lacking [8 and 19]. The current research study findings revealed that the participants who played the mobile game were better able to identify fraudulent websites compared to the participants who read the APWG public education initiative website without any training. This is because the developed mobile game prototype was based on the elements of a theoretical model derived from TTAT. TTAT formulated through an empirical investigation to enhance computer users' phishing threat avoidance behaviour though their motivation. The study incorporated perceived threat, safeguard effectiveness, safeguard cost, self-efficacy, perceived severity and perceived susceptibility elements into the mobile game design context for computer users to thwart phishing attacks. As a result, the current study found that the mobile game is somewhat effective in enhancing the computer user's phishing awareness. Therefore, it is worth understanding how the human aspect of avoiding malicious IT threats can influence in designing security education, for example, when designing security education though web-based training materials.

The research reported in this paper can inform IT security education programs in several aspects. First, it recommends the worth of security awareness, education and training programs. Individuals are more motivated to avoid malicious IT threats and use of safeguarding measures if these security educational programs help them to develop threat perceptions, comprehend the effectiveness of safeguarding measures, lower safeguard costs and increase their self-efficacy. Second, this research suggests that security awareness and training programs should stress both the likelihood of malicious IT threats and the severity of losses caused by the threats. TTAT suggests that emotion-focused coping behaviour is a possible culprit that confuses users' motivation systems at high threat levels. Therefore, public security education initiatives should draw attention to possible emotion-focused coping approaches and help computer users to understand how they unintentionally engage in emotion-focused coping behaviour. Furthermore, individuals should be trained how to stop emotion-focused coping behaviour when they are dealing with malicious IT threats. Such training and education would help computer users to emphasis on the problem-focused coping approach, thus reducing the negative effects of high threat perceptions.

Most organizations do conduct security education, training and awareness programs to regulate employees' malicious IT threat avoidance behaviour and adhere to best security practices against malicious IT threats [30]. However, users at home in a non-work setting are free from those regulations and acting on their own and are easily prey to malicious IT threats. In addition, organizations may consist of either a centralized or decentralized IT security environment [24]. In the centralized organizational IT security environment, security is managed at the enterprise level and employees have no choice to opt out. Those organizations with centralized IT security environment may have sufficient IT infrastructure and strict IT security policies and procedures developed by professional security expertise. In contrast, in the decentralized organizational IT security environment, employees involve in voluntary protective actions such as configuring and enabling their personal firewall/proxy and updating their own antivirus and/or anti-phishing software. These employees in the decentralized IT security environment are highly likely to engage in unsafe computer behaviours and become the weakest link in their organizational security system. Therefore, security education, training and awareness are needed to help these uses to combat against malicious IT threats. The mobile game prototype introduced in this paper as an educational tool focused to teach computer users how to protect themselves against phishing attacks. This is because mobile games can facilitate to embed learning in a natural environment. Therefore, a game based learning approach can contribute to the effectiveness of such IT security programs.

Some organizations such as banks provide help and guidance on their official website to protect their customers against phishing scams. In addition, they provide free anti-virus software such as Kaspersky Internet Security for their customers. However, previous research revealed that reading articles, books, lecture notes or webpages to gain anti-phishing knowledge is still in infancy boring [3, 22 and 23]. On the other hand, how many of computer users are competent enough to properly install, configure and use anti-virus software on their personal computer system is questionable today. Most of them believe that by only installing an anti-virus application does entirely protect their computer system from cyber-attacks. Unfortunately, this is not true. However, organizations like banks should be aware the human element is the weakest link in personal computer use and need to better educate them to combat malicious IT threats. This can be achieved by allowing their customers to free download security educational games (e.g. the mobile game prototype introduced in this paper) for their computers and mobile devices.

## 5. Conclusions and Future Work

Phishing is an online identity theft, which aims to steal sensitive information such as usernames, passwords and online banking details from victims. This research study attempted to evaluate the effectiveness of a mobile game compared to a traditional website in order to protect computer users against phishing attacks. Therefore, a mobile game prototype was developed based on the design introduced by Arachchilage and Cole [3] that aimed to enhance avoidance behaviour through motivation to protect computer users against phishing threats. The APWG public education initiative website was used as a traditional web based learning source. The experiment was conducted through a user study. A think-aloud study was employed along with a pre- and post-test with total 40 participants, where 20 participants were asked to play the mobile game prototype and the other 20 participants were asked to read the website. The study finding revealed that the participants, who played the mobile game, were better able to identify fraudulent websites than the participants, who read the website. We believe that teaching computer users how to prevent from phishing threats using a mobile game, would contribute to enable the cyberspace a secure environment.

However, only five participants who played the game were able to successfully differentiate legitimate websites from phishing websites by looking at URLs in the post-study. The current study identified possible explanations for these results. The mobile game functioned properly, but was still a prototype. It is useful to develop a proper mobile game rather than a prototype with some attractive graphics with visual objects, including more complex URLs and then test on a different sample size to confirm our findings. In addition, limited display size of the mobile phone might have caused a problem for participants especially those who have visual problems. Additionally, it can argue that if participants were allowed to play the game much longer they would have scored more in the post-test. Future research can be conducted with a larger sample size allowing participants to play the game much longer using a high-fidelity mobile game rather than a prototype.

The main objective of our anti-phishing mobile game design was to teach user how to identify phishing URLs, which is one of many ways of identifying a phishing attack. Future research can be conducted on designing a game to teach the other areas such as signs and content of the web page, the lock icons and jargons of the webpage, the context of the email message and the general warning messages displayed on the website.

## 6. References

[1] Amory, A. and Seagram. R. (2003) Educational Game Models: Conceptualization and Evaluation. South African Journal of Higher Education, 17 (2), 206-217.

[2] Anti-Phishing Working Group. (2003) http://www.antiphishing.org/. (06 January 2013)

[3] Arachchilage, N. A. G. and Cole, M. (2011) Design a mobile game for home computer users to prevent from "phishing attacks". Information Society (i-Society), 27-29 June 2011, 485-489, http://ieeexplore.ieee.org/stamp/stamp.jsp?tp=&arnumber=5978543&isnumber=59784. (22 December 2011).

[4] Arachchilage, N. A. G., and Love, S. (2013) A game design framework for avoiding phishing attacks. Computers in Human Behavior, 29(3), 706-714.

[5] Brooke, J. (1996) SUS: A Quick and Dirty Usability Scale, In Usability evaluation in industry, Jordan, P.W., Thomas, B., Weerdmeester, B .A. and McClelland, I.L., (Eds.), Usability Evaluation in Industry, London: Taylor & Francis, http://www.cee.hw.ac.uk/~ph/sus.html., (4 July 2012).

[6] Dhamija, R. and Tygar, J. D. (2005) The battle against phishing: Dynamic Security Skins. Proceedings of the 2005 Symposium on Usable Privacy and Security, Pittsburgh, Pennsylvania, 06 – 08 July 2005, SOUPS '05, ACM Press, New York, NY, 93, 77-88, DOI= http://doi.acm.org/10.1145/1073001.1073009. (20 March 2011).

[7] Downs, J. S., Holbrook, M. and Cranor, L. F. (2007) Behavioural response to phishing risk. Proceedings of the anti-phishing working groups - 2nd annual eCrime researchers summit, October 2007, Pittsburgh, Pennsylvania, 37-44, doi>10.1145/1299015.1299019. (25 March 2011).

[8] Iacovos, K. and Sasse, A. M., (2012) Security Education against Phishing: A Modest Proposal for a Major Rethink, Security & Privacy, IEEE, 10 (2), pp.24-32, March-April 2012, Available at: <doi: 10.1109/MSP.2011.179>, (Accessed 18 July 2012).

[9] Kumaraguru, P., Cranor, L., (2009) Phishguru: a system for educating users about semantic attacks, Carnegie Mellon University, Pittsburgh, PA.

[10] Kumaraguru, P., Rhee, Y., Acquisti, A.,Cranor, L. F., Hong, J. and Nunge, E. (2007) Protecting people from phishing: the design and evaluation of an embedded training email system. Proceedings of the SIGCHI conference on Human Factors in Computing Systems, San Jose, California, USA, April - May 2007.

[11] Kumaraguru, P., Rhee, Y., Sheng, S., Hasan, S., Acquisti, A., Cranor, L. F. and Hong, J. (2007) Getting Users to Pay Attention to Anti-Phishing Education: Evaluation of Retention and Transfer, APWG eCrime Researchers Summit, 4-5October 2007, Pittsburgh, PA, USA.

[12] Kumaraguru, P., Sheng, S., Acquisti, A., Cranor, L. F., and Hong, J. (2007) Teaching Johnny not to fall for phish.Tech.rep., Cranegie Mellon University, http://www.cylab.cmu.edu/files/cmucylab07003.pdf. (12 June 2011).

[13] Liang, H. and Xue, Y. (2010) Understanding Security Behaviours in Personal Computer Usage: A Threat Avoidance Perspective. Association for Information Systems, 11 (7), 394-413.

[14] Ng, B. Y. and Rahim, M. A. (2009) A Socio-Behavioral Study of Home Computer Users Intention to Practice Security, The Ninth Pacific Asia Conference on Information Systems, Bangkok, Thailand.

[15] Parsons, D., Ryu, H. and Cranshaw, M., (2006) A Study of Design Requirements for Mobile Learning Environments, Proceedings of the Sixth IEEE International Conference on Advanced Learning Technologies, 96-100.

[16] Pallant, J. (2007) A step by step guide to data analysis using SPSS for windows (Version15), SPSS survival manual. Buckingham: Open University Press.

[17] PhishTank, (2012) http://www.phishtank.com/. (28 November 2012).

[18] Prensky, M. (2001) Digital Game-Based Learning Revolution. Digital Game-Based Learning, New York.

[19] Purkait, S., (2012) Phishing counter measures and their effectiveness – literature review, Information Management & Computer Security, 20 (5), pp.382 – 420.

[20] Raybourn, E. M. and Waern, A. (2004) Social Learning Through Gaming. Proceedings of CHI 2004, Vienna, Austria, 1733-1734.

[21] Schneier, B. (2000) Semantic Attacks; The Third Wave of Network Attacks, Crypto-Gram Newsletter, October 2000, http://www.schneier.com/crypto-gram-0010.html. (02 April 2011).

[22] Sheng, S., Holbrook, M., Kumaraguru, P., Cranor, L., F. and Downs, J. (2010) Who falls for phish?: a demographic analysis of phishing susceptibility and effectiveness of interventions. 28th international conference on Human factors in computing systems, 10-15 April 2010, Atlanta, Georgia, USA.

[23] Sheng, S., Magnien, B., Kumaraguru, P., Acquisti, A., Cranor, L. F., Hong, J. and Nunge, E. (2007) Anti-Phishing Phil: the design and evaluation of a game that teaches people not to fall for phish. Proceedings of the 3rd symposium on Usable privacy and security, Pittsburgh, Pennsylvania, July 2007.

[24] Warkentin, M. and Johnston, A. C., (2006) IT Security Governance and Centralized Security Controls,in M. Warkentin and R. Vaughn (Eds.) Enterprise Information Assurance and System Security: Managerial and Technical Issues, Hershey, PA: Idea Group Publishing, pp. 16-24.

[25] Woon, I., Tan, G. W. and Low, R. (2005) A Protection Motivation Theory Approach to Home Wireless Security. International Conference on Information Systems, Las Vegas, NV, 367-380.

[26] Workman, M., Bommer, W. H. and Straub, D. (2008) Security Lapses and the Omission of Information Security Measures: A Threat Control Model and Empirical Test. Computers in Human Behavior, 24 (6), 2799-2816.

[27] Ye, Z. and Sean, S. (2002) Trusted Paths for Browsers. Proceedings of the 11th USENIX Security Symposium, USENIX Association, Berkeley, CA, USA, 263 – 279.

[28] Zhang, Y., Egelman, S., Cranor, L. F. and Hong, J. (2007) Phinding phish - Evaluating anti-phishing tools, Proceedings of the 14th Annual Network & Distributed System Security Symposium, 28 February – 2 March 2007, http://lorrie.cranor.org/pubs/toolbars.html, (04 June 2011).

[29] Evers, J., (2007) Security Expert, User education is pointless, Available at: <http://news.com.com/2100-7350_3-6125213.html>, (Accessed 12 June 2011).

[30] Gordon, L. A., M. P. Loeb, W. Lucyshyn, and R. Richardson. (2006) 2006 CSI/FBI computer crime and security survey. Computer Security Institute.